\documentclass[12pt]{article}

\usepackage{hyperref,cite,bm,amsmath,amssymb,a4wide,graphicx,subfigure,wasysym,amsfonts}

\begin{document}

\title{\textbf{Relaxation of the chiral imbalance and the generation of magnetic fields in magnetars}}

\author{Maxim Dvornikov$^{a,b,c}$\thanks{maxdvo@izmiran.ru}
\\
$^{a}$\small{\ Pushkov Institute of Terrestrial Magnetism, Ionosphere} \\
\small{and Radiowave Propagation (IZMIRAN),} \\
\small{108840 Moscow, Troitsk, Russia;} \\
$^{b}$\small{\ Physics Faculty, National Research Tomsk State University,} \\
\small{36 Lenin Avenue, 634050 Tomsk, Russia;} \\
$^{c}$\small{\ II. Institute for Theoretical Physics, University of Hamburg,} \\ 
\small{149 Luruper Chaussee, D-22761 Hamburg, Germany}}

\date{}

\maketitle

\begin{abstract}
The model for the generation of magnetic fields in a neutron star, based on the magnetic field instability caused by the electroweak interaction between electrons and nucleons, is developed. Using the methods of the quantum field theory, the helicity flip rate of electrons in their scattering off protons in dense matter of a neutron star is calculated. The influence of the electroweak interaction between electrons and background nucleons on the process of the helicity flip is studied. The kinetic equation for the evolution of the chiral imbalance is derived. The obtained results are applied for the description of the magnetic fields evolution in magnetars.
\end{abstract}

\section{Introduction}

Some neutron stars (NS) can possess extremely strong magnetic fields
$B\gtrsim10^{15}\thinspace\text{G}$. These NSs are called magnetars~\cite{MerPonMel15}.
Despite a long observational history of magnetars and numerous theoretical
models for the generation of their magnetic fields, nowadays there
is no commonly accepted mechanism explaining the origin of magnetic
fields in these compact stars.

While constructing a model for the magnetic field in a magnetar one encounters the following main difficulties. Firstly,
it is necessary to explain the generation of magnetic fields which are rather strong $B\gtrsim10^{15}\thinspace\text{G}$
and large scale $\Lambda_{\mathrm{B}}\sim R_{\mathrm{NS}}$,
where $R_{\mathrm{NS}}\sim10\thinspace\text{km}$ is the NS radius. Some popular scenarios in Refs.~\cite{DunTho92,VinKui06}
predict the creation of such magnetic fields during several seconds after the supernova (SN) collapse. However these models require
quite peculiar initial conditions. Secondly, it is unclear why the
magnetic field, generated during such a short time interval, should be confined inside NS for $t\gtrsim10^{3}\thinspace\text{yr}$,
which is the typical age of young magnetars, and only afterwards be
released outside NS to produce the gamma or X-ray radiation
of magnetars~\cite{MerPonMel15}. The model, proposed in Ref.~\cite{ThoLyuKul02} to explain the electromagnetic radiation
of magnetars, based on the magnetic
field release through the cracks in the NS crust is likely to
be rather catastrophic.

Recently several attempts were made in Refs.~\cite{SigLei16,Yam15}
to solve
the problem of the magnetic field generation in magnetars using the chiral magnetic effect~\cite{MirSho15}.
This effect was also used in Ref.~\cite{ChaZhi10} for the generation of toroidal magnetic fields in NSs. Another mechanism to explain the creation
of strong cosmic magnetic fields based on the magnetic field instability
driven by the parity violating interaction was proposed in Refs.~\cite{Vil80,Rub86}. Recently this idea was revised in Ref.~\cite{BoyRucSha12}.

In Refs.~\cite{DvoSem15a,DvoSem15b,DvoSem15c} we developed the new
model for the generation of magnetic fields in magnetars. The main
mechanism, underlying our model, is the amplification of a seed magnetic field
due to the field instability in nuclear matter driven by the electron-nucleon
($eN$) electroweak interaction. In frames of our approach, we obtained
the amplification of the magnetic field from $B_{0}=10^{12}\thinspace\text{G}$, which is a typical field strength in a young pulsar, to the values predicted
in magnetars. The length scale of the magnetic field generated was
comparable with the NS radius. The magnetic field was amplified
in the time interval $(10^{3}-10^{5})\thinspace\text{yr}$, depending
on its length scale. Besides the creation of strong magnetic fields,
we also predict the generation of the magnetic helicity in magnetars in frames of our model~\cite{DvoSem15b}.

Despite the explanation of various properties of magnetars in Refs.~\cite{DvoSem15a,DvoSem15b,DvoSem15c},
some of the features of this model should be substantiated by more detailed calculations based on reliable methods of the quantum field theory (QFT).
This work is devoted to the further development
of the proposed description of the magnetic fields generation in magnetars.

The present paper is organized in the following way. In Sec.~\ref{sec:MOD},
we recall the main features of the model for the magnetic fields generation in magnetars.
In Sec.~\ref{sec:COLLISIONS},
we study electron-proton ($ep$) collisions in dense matter of NS. In particular,
in Sec.~\ref{sub:MAT}, we compute the total probability of the helicity flip of an electron in an $ep$ collision.
Then, in Sec.~\ref{sub:KIN}, we derive the kinetic equation for the chiral imbalance. The relaxation
of the chiral imbalance from the point of view of the thermodynamics is studied in Sec.~\ref{sub:TERM}.
The obtained results are applied in Sec.~\ref{sec:EVOL} for the description of the magnetic field
generation in magnetars. We summarize
and discuss our results in Sec.~\ref{sec:CONCL}. The solution of
the Dirac equation for a massive electron electroweakly interacting
with background nucleons is provided in Appendix~\ref{sec:SOLDIREQ}. Some details of the computation
of the intergals over the phase space are given in Appendix~\ref{sec:INTCALC}. In Appendix~\ref{sec:KINEQ}, we
derive the kinetic equations for the occupation numbers of relativistic electrons. The energy balance in a magnetar
is discussed in Appendix~\ref{sec:ENSOR}.

\section{The model for the magnetic fields generation in magnetars\label{sec:MOD}}

In this section we briefly describe the model for the generation
of strong large-scale magnetic fields in magnetars based on the instability
of the magnetic field driven by the parity violating $eN$ electroweak interaction.

The dense matter of NS is known to consist of ultrarelativistic electrons
and nonrelativistic nucleons, which are neutrons and protons. This matter is
supposed to have zero macroscopic velocity and polarization. In this matter, electrons
interact with nucleons by the parity violating electroweak
forces. We found in Refs.~\cite{DvoSem15a,DvoSem15b} that, in the
external magnetic field $\mathbf{B}$, there is the induced anomalous
electric current of electrons $\mathbf{J}$, which has the form,
\begin{equation}\label{eq:JCS}
  \mathbf{J}=\Pi\mathbf{B},
  \quad
  \Pi=\frac{2\alpha_{\mathrm{em}}}{\pi}
  \left(
    \mu_{5}+V_{5}
  \right),
\end{equation}
where $\alpha_{\mathrm{em}}\approx7.3\times10^{-3}$ is the fine structure
constant, $\mu_{5}=\left(\mu_{\mathrm{R}}-\mu_{\mathrm{L}}\right)/2$
is the chiral imbalance, $\mu_{\mathrm{R},\mathrm{L}}$ are the chemical
potentials of the right and left electrons, $V_{5}=\left(V_{\mathrm{L}}-V_{\mathrm{R}}\right)/2\approx G_{\mathrm{F}}n_{n}/2\sqrt{2}$,
$V_{\mathrm{L},\mathrm{R}}$ are the effective potentials of the interaction
of left and right electrons with background nucleons (mainly with
neutrons), $G_{\mathrm{F}}\approx1.17\times10^{-5}\thinspace\text{GeV}^{-2}$
is the Fermi constant, and $n_{n}$ is the neutron density. The explicit
values of $V_{\mathrm{L,R}}$ are given in Eq.~(\ref{eq:VLR}). The
current in Eq.~(\ref{eq:JCS}) was obtained in Refs.~\cite{DvoSem15a,DvoSem15b}
on the basis of the exact solution of the Dirac equation for an ultrarelativistic
electron interacting with a background matter under the influence an an external magnetic field.
This
current is additive to the ohmic current $\mathbf{J}_{\mathrm{ohm}}=\sigma_{\mathrm{cond}}\mathbf{E}$,
where $\sigma_{\mathrm{cond}}$ is the matter conductivity and $\mathbf{E}$
is the electric field.

Basing on Eq.~(\ref{eq:JCS}), in Ref.~\cite{DvoSem15b} we derived
the system of the evolution equations for the spectrum of the helicity
density $h(k,t)$, the spectrum of the magnetic energy density $\rho_{\mathrm{B}}(k,t)$,
and the chiral imbalance, which reads
\begin{align}
  \frac{\partial h(k,t)}{\partial t} = &
  -\frac{2k^{2}}{\sigma_{\mathrm{cond}}}h(k,t) +
  \frac{8\alpha_{\mathrm{em}}
  \left[
    \mu_{5}(t)+V_{5}
  \right]}
  {\pi\sigma_{\mathrm{cond}}}\rho_{\mathrm{B}}(k,t),
  \label{eq:heq}
  \\
  \frac{\partial\rho_{\mathrm{B}}(k,t)}{\partial t}= &
  - \frac{2k^{2}}{\sigma_{\mathrm{cond}}}\rho_{\mathrm{B}}(k,t) +
  \frac{2\alpha_{\mathrm{em}}
  \left[
    \mu_{5}(t)+V_{5}
  \right]}
  {\pi\sigma_{\mathrm{cond}}}k^{2}h(k,t),
  \label{eq:rhoeq}
  \\
  \frac{\mathrm{d}\mu_{5}(t)}{\mathrm{d}t} = &
  \frac{\pi\alpha_{\mathrm{em}}}{\mu_{e}^{2}\sigma_{\mathrm{cond}}}
  \int\mathrm{d}k
  \left\{
    k^{2}h(k,t) -
    \frac{4\alpha_{\mathrm{em}}}{\pi}
    \left[
      \mu_{5}(t)+V_{5}
    \right]
    \rho_{\mathrm{B}}(k,t)
  \right\} -
  \Gamma_{f}\mu_{5}(t),
  \label{eq:mu5eq}
\end{align}
where $\Gamma_{f}$ is the helicity flip in $ep$ collisions (see Sec.~\ref{sec:COLLISIONS}) and
$\mu_{e}$ is the mean chemical potential of the electron gas. The
functions $h(k,t)$ and $\rho_{\mathrm{B}}(k,t)$ are related to the
magnetic helicity $H(t)$ and the strength of the magnetic field by
\begin{equation}\label{eq:hdef}
  H(t) = V \int h(k,t) \mathrm{d}k,
  \quad
  \frac{1}{2}B^{2}(t) = \int \rho_{\mathrm{B}}(k,t) \mathrm{d}k,
\end{equation}
where $V$ is the normalization volume. The integration in Eq.~(\ref{eq:hdef}) is over
all the range of the wave number $k$ variation. We also mention that
in Eq.~(\ref{eq:hdef}) we assume the isotropic spectra.

Eqs.~(\ref{eq:heq}) and~(\ref{eq:rhoeq}) for $h(k,t)$ and $\rho_{\mathrm{B}}(k,t)$
are the direct consequence of the modified Faraday equation (see Eq.~\eqref{FE} in Sec.~\ref{sec:EVOL}) completed
by the anomalous current in Eq.~(\ref{eq:JCS}). The first two terms
in Eq.~(\ref{eq:mu5eq}) for $\mu_{5}(t)$ result from Eq.~(\ref{eq:heq})
and the conservation law,
\begin{equation}\label{eq:consnLR}
  \frac{\mathrm{d}}{\mathrm{d}t}
  \left(
    n_{\mathrm{R}}-n_{\mathrm{L}} +
    \frac{\alpha_{\mathrm{em}}}{\pi V}H
  \right)=0,
\end{equation}
where $n_{\mathrm{R,L}}$ are the number densities of right and left
electrons. Note that Eq.~(\ref{eq:consnLR}) is a consequence of
the Adler anomaly for ultrarelativistic electrons~\cite[p.~359--420]{Adl69}.

The last term in rhs of Eq.~(\ref{eq:mu5eq}), $\Gamma_{f}\mu_{5}$, was
accounted for phenomenologically. It is based on the fact that the electron
helicity is changed in an $ep$ collision. Typically electrons
are ultrarelativistic in NS. However they have a nonzero mass. Thus,
in Ref.~\cite{DvoSem15a}, we estimated $\Gamma_{f}$ as
\begin{equation}\label{eq:Gfsimp}
  \Gamma_{f} \sim
  \left(
    \frac{m}{\mu_{e}}
  \right)^{2}
  \nu_{\mathrm{coll}} \sim
  \left(
    \frac{m}{\mu_{e}}
  \right)^{2}
  \frac{\omega_{p}^{2}}{\sigma_{\mathrm{cond}}},
\end{equation}
where $m$ is the electron mass, $\nu_{\mathrm{coll}}$ is the frequency
of $ep$ collisions, and $\omega_{p}$ is the plasma frequency in
the degenerate plasma. Eq.~(\ref{eq:Gfsimp}) is based on the relation
between $\nu_{\mathrm{coll}}$ and $\sigma_{\mathrm{cond}}$ in the
classical Lorentz plasma~\cite[pp.~66--67]{AleBogRuk84}.

Therefore, to complete the theoretical substantiation of the main equations of the model in Refs.~\cite{DvoSem15a,DvoSem15b,DvoSem15c} it is necessary
to consider the helicity flip of electrons in $ep$ collisions in dense matter of NS using the QFT methods. Moreover it is interesting
to examine the influence of the electroweak interaction between electrons and nucleons on this process.

\section{Electron-proton collisions in dense plasma\label{sec:COLLISIONS}}

In this section we shall study the helicity flip of electrons scattering
off protons in nuclear matter consisting of degenerate neutrons, protons,
and electrons. Note that, while studying the scattering process, we
shall exactly take into account the electroweak interaction between
electrons and nucleons. As a result, in Sec.~\ref{sub:MAT}, we find the
helicity flip rate of electrons in the considered matter. Then, in Sec.~\ref{sub:KIN},
we derive the the kinetic equation for the chiral imbalance. Finally, in Sec.~\ref{sub:TERM},
we analyze the chiral imbalance evolution from the point of view of thermodynamics.

In NS, the helicity of a massive electron can be changed in $ep$ and
$ee$ scatterings owing to the electromagnetic interaction mediated by the virtual plasmon exchange,
as well as in the interaction of an electron with the anomalous magnetic moment of a neutron. As
found in Ref.~\cite{Kel73}, the rate of $ep$ reactions in dense
matter of NS is higher than that of the others. Therefore, in our analysis,
we shall account for only $ep$ collisions.

\subsection{Helicity flip rate in $ep$ collisions\label{sub:MAT}}

\begin{figure}
  \centering
  \includegraphics[scale=.07]{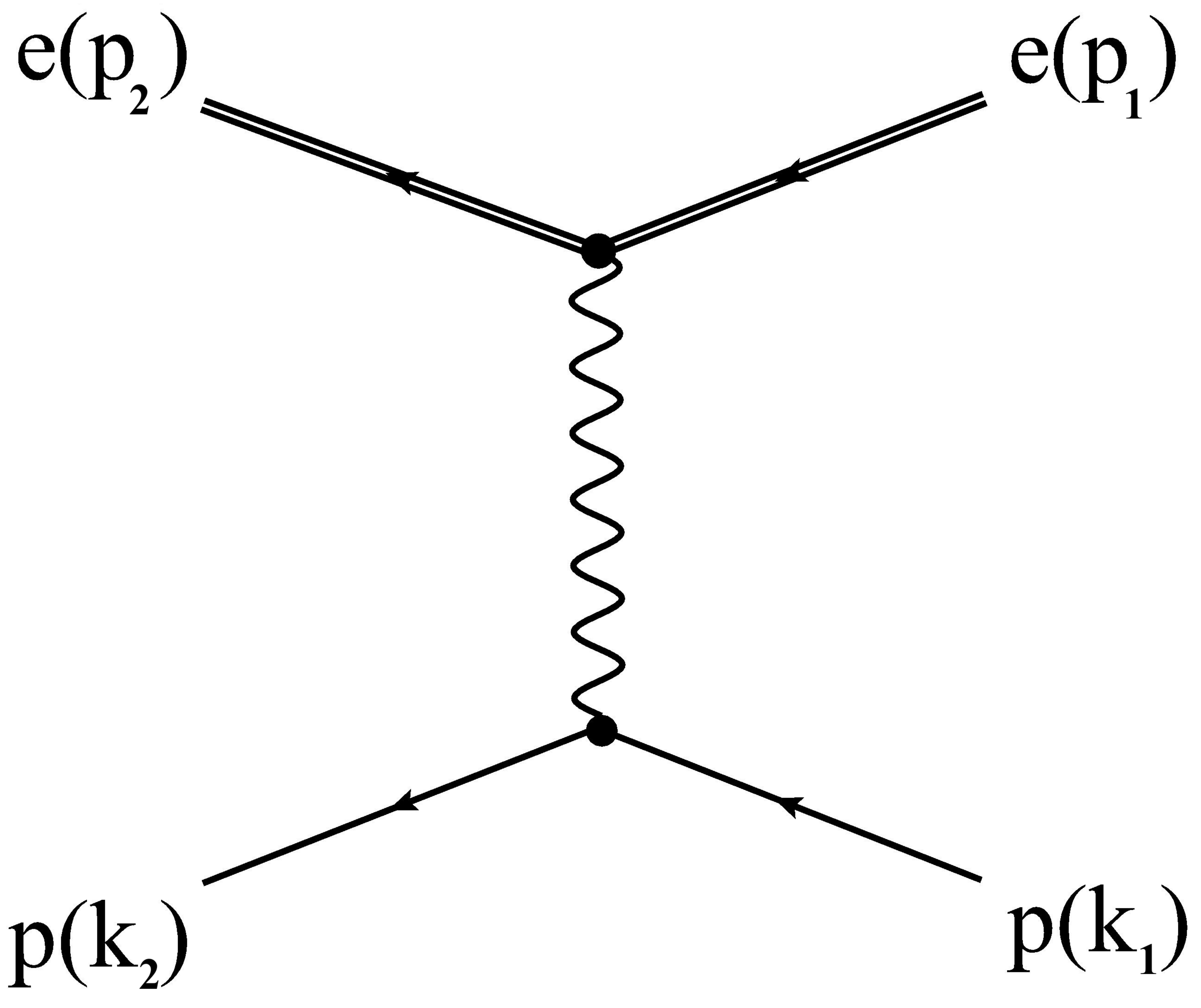}
  \protect\caption{The Feynman diagram for the $ep$ scattering.
  Broad electron lines mean the basis spinors corresponding to the 
  exact solution of the Dirac equation in Eq.~\eqref{eq:upm}.\label{fig:FD}}
\end{figure}

The matrix element for the $ep$ collision, due to the electromagnetic
interaction, has the form,
\begin{equation}\label{eq:matrel}
  \mathcal{M} =
  \frac{\mathrm{i}e^{2}}{
  \left(
    k_{1}-k_{2}
  \right)^{2}}
  \bar{u}_e(p_{2})\gamma^{\mu}u_e(p_{1}) \cdot
  \bar{u}_p(k_{2})\gamma_{\mu}u_p(k_{1}),
\end{equation}
where $e>0$ is the absolute value of the electron charge, $\gamma^{\mu} = \left( \gamma^{0},\bm{\gamma} \right)$
are the Dirac matrices, $u_{e,p}$ are the bispinors corresponding to the wave functions of electrons and protons,
$p_{1,2}^{\mu} = \left( E_{1,2},\mathbf{p}_{1,2} \right)$
and $k_{1,2}^{\mu} = \left( \mathcal{E}_{1,2},\mathbf{k}_{1,2} \right)$
are the four momenta of electrons and protons. The momenta of incoming
particles are marked with the label 1 and that of the outgoing particles
with the label 2. The Feynman diagram for this process is shown in Fig.~\ref{fig:FD}.

Besides the plasmon exchange with a proton, an electron can electroweakly interact with nucleons in NS.
To account for this interaction in the matrix element in Eq.~\eqref{eq:matrel},
we should use the spinors corresponding to the exact solutions of the Dirac
equation for an electron interacting with a background matter found
in Appendix~\ref{sec:SOLDIREQ} instead of the solutions of the Dirac equation in vacuum.

We shall be interested in the reactions where the electron helicity flips.
We shall start with the analysis of $e_\mathrm{R} \to e_\mathrm{L}$ transitions. According
to Eq.~(\ref{eq:matrel}) it is necessary to compute the following
quantity:
\begin{equation}
  J^{\mu} =
  \left(
    J_{0},\bm{J}
  \right) =
  \bar{u}_{-}(p_{2})\gamma^{\mu}u_{+}(p_{1}).
\end{equation}
Using Eqs.~(\ref{eq:upm}) and~(\ref{eq:normcoef}) one gets
\begin{align}\label{eq:J0J}
  J_{0}= & -\frac{mP_{0}
  \left[
    p_{1}+p_{2}+E_{+}(p_{1})+E_{-}(p_{2})-2\bar{V}
  \right]}{
  2\sqrt{E_{0+}(p_{1})E_{0-}(p_{2})
  \left[
    E_{-}(p_{2})+p_{2}-V_{\mathrm{R}}
  \right]
  \left[
    E_{+}(p_{1})+p_{1}-V_{\mathrm{L}}
  \right]}},
  \nonumber
  \\
  \bm{J} = & -\frac{m\mathbf{P}
  \left[
    p_{1}-p_{2}+E_{+}(p_{1})-E_{-}(p_{2})-2V_{5}
  \right]}{
  2\sqrt{E_{0+}(p_{1})E_{0-}(p_{2})
  \left[
    E_{-}(p_{2})+p_{2}-V_{\mathrm{R}}
  \right]
  \left[
    E_{+}(p_{1})+p_{1}-V_{\mathrm{L}}
  \right]}},
\end{align}
where
\begin{equation}\label{eq:P0P}
  P_{0} = w_{-}^{\dagger}(\mathbf{p}_{2})w_{+}(\mathbf{p}_{1}),
  \quad
  \mathbf{P} =
  w_{-}^{\dagger}(\mathbf{p}_{2}) \bm {\sigma}w_{+}(\mathbf{p}_{1}).
\end{equation}
Here $\bm{\sigma}$ are the Pauli matrices and $\bar{V}=\left(V_{\mathrm{L}}+V_{\mathrm{R}}\right)/2$.
To obtain Eq.~(\ref{eq:J0J}) we use the Dirac matrices in the chiral
representation~\cite[pp.~691--696]{ItzZub80}.

As shown in Ref.~\cite[pp.~205--209]{LifPit81}, while considering collisions in plasma owing to the long range Coulomb forces,
one should use the approximation of the elastic scattering. Thus, studying $R \to L$ transitions, we should take that $E_{+}(p_{1})=E_{-}(p_{2})$.
Considering ultrarelativistic electrons and using the fact that $E_{\pm}(p_{1,2}) = p_{1,2} + V_\mathrm{R,L}$, see Eq.~\eqref{eq:enlev},
this condition is equivalent to $p_{1}-p_{2}=2V_{5}$. On the basis of Eq.~\eqref{eq:J0J} we get that $\bm{J}=0$. To compute $P_{0}$ in Eq.~(\ref{eq:P0P})
we use the explicit form of the helicity amplitudes in Eq.~(\ref{eq:helampl}). The direct calculation shows that
\begin{equation}\label{eq:P02}
  |P_{0}|^{2} = \frac{1}{2}
  \left[
    1 -
    \left(
      \mathbf{n}_{1}\cdot\mathbf{n}_{2}
    \right)
  \right],
\end{equation}
where $\mathbf{n}_{1,2}$ are the unit vectors along $\mathbf{p}_{1,2}$.

Thus, using Eqs.~(\ref{eq:J0J}) and~(\ref{eq:P02}), one gets that the square of matrix element in Eq.~(\ref{eq:matrel})
has the form,
\begin{equation}\label{eq:M2ew}
  |\mathcal{M}|^{2} = 
  e^{4}m^{2}
  \frac{
  \left(
    p_{1}+p_{2}
  \right)^{2}
  \left[
    1-
    \left(
      \mathbf{n}_{1}\cdot\mathbf{n}_{2}
    \right)
  \right]}{
  8
  \left(
    p_{1}-V_{5}
  \right)^{2}
  \left(
    p_{2}+V_{5}
  \right)^{2}}
  \frac{
  \left[
    \mathcal{E}_{1}\mathcal{E}_{2}+M^{2}+
    \left(
      \mathbf{k}_{1}\cdot\mathbf{k}_{2}
    \right)
  \right]}{
  \left[
    \left(
      \mathcal{E}_{1}-\mathcal{E}_{2}
    \right)^{2} -
    \left(
      \mathbf{k}_{1}-\mathbf{k}_{2}
    \right)^{2}
  \right]^{2}},
\end{equation}
where we keep only the leading term in the electron mass. Note that, the contribution of protons, which are considered
to be unpolarized, to $|\mathcal{M}|^{2}$ can be found using the standard methods
(see, e.g., Ref.~\cite[pp.~252--256]{BerLifPit82}).

The total probability of the process has the form~\cite[pp.~248--249]{BerLifPit82},
\begin{align}\label{eq:Wew}
  W= & \frac{V}{2(2\pi)^{8}}
  \int
  \frac{\mathrm{d}^{3}p_{1}\mathrm{d}^{3}p_{2}
  \mathrm{d}^{3}k_{1}\mathrm{d}^{3}k_{2}}
  {\mathcal{E}_{1}\mathcal{E}_{2}}
  \delta^{4}
  \left(
    p_{1}+k_{1}-p_{2}-k_{2}
  \right)
  |\mathcal{M}|^{2}
  \nonumber
  \\
  & \times
  f_{e}(E_{1}^+ -\mu_\mathrm{R})
  \left[
    1-f_{e}(E_{2}^- -\mu_\mathrm{L})
  \right]
  f_{p}(\mathcal{E}_{1}-\mu_{p})
  \left[
    1-f_{p}(\mathcal{E}_{2}-\mu_{p})
  \right],
\end{align}
where we sum over the polarizations of outgoing protons. Here
$f_{e,p}(E) = \left[ \exp(\beta E)+1 \right]^{-1}$ are the Fermi-Dirac distributions
of electrons and protons, $\beta=1/T$ is the reciprocal temperature,
$\mu_{p}$ is the chemical potential of protons, and $V$ is the normalization
volume. In Eq.~(\ref{eq:Wew}) we assume that incoming and
outgoing electrons have different chemical potentials: $\mu_\mathrm{R}$
and $\mu_\mathrm{L}$ respectively. Protons and electrons are taken to be
in the thermal equilibrium having the same temperature $T$. The expression
for the total probability corresponds to the normalization of the
electron wave functions in Eq.~\eqref{eq:norm}.

Since we are looking for the probability of the  $R\to L$ process in the leading order in
the electron mass $m$, and $|\mathcal{M}|^{2}\sim m^{2}$ in Eq.~(\ref{eq:M2ew}),
we can take that electrons are massless in the computation of the
integral over the phase space in Eq.~\eqref{eq:Wew}. Moreover we assume that the electron
gas is highly degenerate, that leads to $f_{e}(E_{1}-\mu_{1})=\theta\left(\mu_{1}-E_{1}\right)$
and $1-f_{e}(E_{2}-\mu_{2})=\theta\left(E_{2}-\mu_{2}\right)$, where
$\theta(z)$ is the Heaviside step function.

The direct standard computation of the integrals over the momenta of electrons and protons in Eq.~\eqref{eq:Wew} gives
(see Appendix~\ref{sec:INTCALC}),
\begin{equation}\label{eq:Wfin}
  W(R \to L) = W_{0}
  \left(
    \mu_\mathrm{R}-\mu_\mathrm{L}
  \right)
  \theta
  \left(
    \mu_\mathrm{R}-\mu_\mathrm{L}
  \right),
  \quad
  W_{0} = \frac{Ve^{4}}{32\pi^{5}}
  \frac{m^{2}M}{\mu_{e}}T
  \left[
    \ln
    \left(
      \frac{48\pi}{\alpha_{\mathrm{em}}}
    \right)-4
  \right],
\end{equation}
where $M$ is the proton mass. Note that, while deriving Eq.~\eqref{eq:Wfin}, we exactly account for the dependence on potentials of the 
interaction of electrons with matter $V_\mathrm{L,R}$. If we study $L\to R$ transitions, the calculation of the total probability
is analogous to the $R\to L$ case. One can show that the expression for $W(L \to R)$, in this case
coincides with that in Eq.~(\ref{eq:Wfin}), where we should replace $\mu_\mathrm{R}\leftrightarrow\mu_\mathrm{L}$. For the sake of brevity
we omit these computations.

One of the most important results is the dependence of $W$ in Eq.~\eqref{eq:Wfin} on the chemical potentials: $W(R \to L) \sim \left(
\mu_\mathrm{R}-\mu_\mathrm{L} \right)$. Note that such a feature is independent of the assumption of the elasticity of $ep$ collisions, which was made while deriving Eq.~\eqref{eq:Wfin}.
This dependence of $W$ results from the expression for the energy of ultrarelativistic electrons in matter in Eq.~\eqref{eq:enlev},
$E_{1,2}^\pm = p_{1,2} + V_\mathrm{R,L}$, which should be taken into account both in the energy conservation delta function,
$\delta(E_{1}^+ + \mathcal{E}_1 - E_{2}^- - \mathcal{E}_2)$, and in the energy distribution functions of electrons. It leads to
the fact that the potentials $V_\mathrm{L,R}$ do not contribute to the factor $\left( \mu_\mathrm{R}-\mu_\mathrm{L} \right)$ in Eq.~\eqref{eq:Wfin}.
If inelastic effects are accounted for, there can be a dependence of the function $W_0$ on $V_\mathrm{L,R}$.

Protons are taken to be nonrelativistic and unpolarized. If we introduce the analogues of $V_\mathrm{L,R}$ for protons, see Eq.~~\eqref{eq:VLR},
and denote them as
$V_\mathrm{L,R}^{(p)}$, then, sing Eq.~\eqref{eq:enlev}, we can estimate the contribution of the electroweak interaction to the proton energies as
$\Delta(\mathcal{E}_{1,2})_\mathrm{EW} \sim \bar{V}_p \mp k_{1,2} V_5^{(p)}/M$, where $\bar{V}_p = [V_\mathrm{L}^{(p)} + V_\mathrm{R}^{(p)}]/2$.
Thus one can see that, in the energy conservation delta function in Eq.~\eqref{eq:Wew}:
$|\Delta(\mathcal{E}_1 - \mathcal{E}_2)_\mathrm{EW}| \lesssim p_\mathrm{F_p} V_5^{(p)}/M \sim 0.1 V_5 \ll V_5$, since the Fermi momentum
of protons in NS is $p_\mathrm{F_p} \sim 10^2\,\text{MeV}$, as well as $M\sim 1\,\text{GeV}$ and $V_5^{(p)} \sim V_5$. As was mentioned above, for electrons we have
$|\Delta({E}_1 - E_2)_\mathrm{EW}| = 2 V_5$. Thus the contribution of the electroweak interaction of protons to the conservation of energy is negligible
compared to that of the $eN$ interaction: $|\Delta(\mathcal{E}_1 - \mathcal{E}_2)_\mathrm{EW}| \ll |\Delta({E}_1 - E_2)_\mathrm{EW}|$.

\subsection{Kinetics of the chiral imbalance\label{sub:KIN}}

Basing on Eq.~(\ref{eq:Wfin}) and analogous expression for $L\to R$ transitions, one gets the kinetic equations for
the total numbers of right and left electrons $N_{\mathrm{R},\mathrm{L}}$ as
\begin{align}\label{eq:kineqN}
  \frac{\mathrm{d}N_{\mathrm{R}}}{\mathrm{d}t} = &
  - W(R \to L) + W(L \to R) =
  -W_{0}
  \left(
    \mu_{\mathrm{R}}-\mu_{\mathrm{L}}
  \right),
  \notag
  \\
  \frac{\mathrm{d}N_{\mathrm{L}}}{\mathrm{d}t} = &
  - W(L \to R) + W(R \to L) =
  -W_{0}
  \left(
    \mu_{\mathrm{L}}-\mu_{\mathrm{R}}
  \right),
\end{align}
Note that one can also derive Eq.~\eqref{eq:kineqN} from the Boltzmann kinetic equation for the distribution functions of right and left electrons
accounting for the collision integrals describing the interaction with protons (see Eqs.~\eqref{eq:kinNLR} and~\eqref{eq:WLR} in Appendix~\ref{sec:KINEQ}).

Defining the number densities of left and right electrons $n_{\mathrm{R},\mathrm{L}}=N_{\mathrm{R},\mathrm{L}}/V$ and using the expression for $n_{\mathrm{R},\mathrm{L}}$
in terms of the distribution function,
\begin{equation}\label{eq:nmuV}
  n_{\mathrm{R},\mathrm{L}} =
  2\int\frac{\mathrm{d}^{3}p}{(2\pi)^{3}}
  \frac{1}{\exp
  \left[
  \beta
    \left(
      p+V_{\mathrm{R},\mathrm{L}}-\mu_{\mathrm{R},\mathrm{L}}
    \right)
  \right] + 1}
  \approx\frac{
  \left(
    \mu_{\mathrm{R},\mathrm{L}}-V_{\mathrm{R},\mathrm{L}}
  \right)^{3}}
  {3\pi^{2}},
\end{equation}
we get that $\mathrm{d}\left(n_{\mathrm{R}}-n_{\mathrm{L}}\right)/\mathrm{d}t\approx2\dot{\mu}_{5}\mu_{e}^{2}/\pi^{2}$,
where it is accounted for that $\dot{V}_5 = 0$ and $\mu_{5}\ll\mu_{e}$. Finally one can derive the kinetic equation for $\mu_{5}$,
\begin{equation}\label{eq:m5kincorr}
  \frac{\mathrm{d}\mu_{5}}{\mathrm{d}t} = -\Gamma_{f}\mu_{5},  
  \quad
  \Gamma_{f} = \frac{\alpha_{\mathrm{em}}^{2}}{\pi}
  \left[
    \ln
    \left(
      \frac{48\pi}{\alpha_{\mathrm{em}}}
    \right) - 4
  \right]
  \left(
    \frac{m}{\mu_{e}}
  \right)^{2}
  \left(
    \frac{M}{\mu_{e}}
  \right)T,
\end{equation}
where we use Eqs.~(\ref{eq:Wfin}) and~(\ref{eq:kineqN}).

It is necessary to mention that the value of $\Gamma_{f}$ in Eq.~(\ref{eq:Gfsimp}) is different from that 
used in Refs.~\cite{DvoSem15a,DvoSem15b,DvoSem15c}. The reason for
the discrepancy between $\Gamma_{f}$ in Eqs.~(\ref{eq:Gfsimp})
and~(\ref{eq:m5kincorr}) consists in the fact that in Ref.~\cite{DvoSem15a}
we relied on the results of Ref.~\cite{Kel73}, where the scattering
of unpolarized electrons off protons was studied. However,
in our case it is essential to have the fixed opposite polarizations of incoming
and outgoing electrons. This fact explains,
e.g., that $\Gamma_{f}$ in Eq.~(\ref{eq:m5kincorr}) is linear in
$T$ whereas that in Eq.~(\ref{eq:Gfsimp}) is proportional to $T^{2}$.

In our study of the chiral imbalance evolution we do not account for
the influence of the magnetic field present in Eqs.~(\ref{eq:heq})-(\ref{eq:mu5eq}).
In particular, we derive the kinetic equations in the leading nonzero order in $\alpha_{\mathrm{em}}$.
If we used the exact solutions of the Dirac equation for an electron
interacting with background matter and an external magnetic field~\cite{DvoSem15a,DvoSem15b}
in the calculation of the matrix element in Eq.~(\ref{eq:matrel}),
it would give a higher order correction in $\alpha_{\mathrm{em}}$
to $\Gamma_{f}$ in Eq.~(\ref{eq:m5kincorr}).

We also mention that $\Gamma_{f}$ was recently calculated in Ref.~\cite{GraKapRed15}.
The value of $\Gamma_{f}$ obtained in Ref.~\cite{GraKapRed15} is
independent of $T$ since it was assumed that protons are nondegenerate.
This assumption is valid when the early stages of the NS evolution are
considered. In the present work, we study the magnetic field generation in a thermally
relaxed NS at $t\gtrsim10^{2}\thinspace\text{yr}$ after the onset
of the SN collapse (see Sec.~\ref{sec:EVOL} below). At this time, the
proton component of the NS matter should be taken as degenerate.
The Fermi temperature for protons, which are nonrelativistic, can be estimated as
$T_\mathrm{deg} \sim n_p^{2/3} / M$, where $n_p$ is the number density of protons.
Assuming that $n_p = 9 \times 10^{36}\,\text{cm}^{-3}$
(see, e.g., Ref.~\cite{DvoSem15b} and Sec.~\ref{sec:EVOL}), we get that $T_\mathrm{deg} \sim 10^{10}\,\text{K}$.
Below, in Sec.~\ref{sec:EVOL} we assume that the initial NS temperature is $T_0 \leq 10^9\,\text{K}$,
i.e. $T_0 \ll T_\mathrm{deg}$. Therefore, in our model protons are degenerate with the high level of accuracy.
Note that $\Gamma_{f} \sim \alpha^2_{\mathrm{em}}$ in Eq.~\eqref{eq:m5kincorr} as in Ref.~\cite{GraKapRed15}.

\subsection{Thermodynamic description of the chiral imbalance relaxation\label{sub:TERM}}

Recently in Ref.~\cite{SigLei16} it was suggested that the kinetics of the chiral imbalance in the system of left and right electrons,
electroweakly interacting with matter, obeys the equation,
\begin{equation}\label{eq:siglwrong}
  \frac{\mathrm{d}\mu_5}{\mathrm{d} t} = - \Gamma_f (\mu_5 + V_5),
\end{equation}
rather than Eq.~\eqref{eq:m5kincorr}, which results from our calculations. Nevertheless, it is possible to show that Eq.~\eqref{eq:siglwrong} contradicts
the laws of thermodynamics.

Using Eq.~\eqref{eq:nmuV}, one can rewrite Eq.~\eqref{eq:siglwrong} in the form,
\begin{equation}\label{eq:siglwrongn5}
  \frac{\mathrm{d}}{\mathrm{d} t}
  \left(
    n_{\mathrm{R}} - n_{\mathrm{L}}
  \right) = - \frac{2\mu_e^2}{\pi^2}\Gamma_f (\mu_5 + V_5).
\end{equation} 
One can see in Eq.~\eqref{eq:siglwrongn5} that the state of equilibrium, in which $n_{\mathrm{R,L}} = \text{const}$, would be achieved at
$\tilde{\mu}_{\mathrm{R}} = \tilde{\mu}_{\mathrm{L}}$, where $\tilde{\mu}_{\mathrm{L,R}} = \mu_{\mathrm{L,R}} - V_{\mathrm{L,R}}$, rather than at
${\mu}_{\mathrm{R}} = {\mu}_{\mathrm{L}}$, as it is required by the laws of thermodynamics~\cite[p.~306]{LanLif02}. Note that the quantities
$\tilde{\mu}_{\mathrm{L,R}} = \tilde{\mu}_{\mathrm{L,R}}(P,T)$, where $P$ is the pressure in the system, which are introduced formally,
are the chemical potentials at the absence of the background matter. 

The analysis of the state of equilibrium in the system of left and right electrons is a particular example of the description of the equilibrium of
a body in the external field $\mathcal{V}(\mathbf{r})$. As shown in Ref.~\cite[pp.~73--74]{LanLif02}, the equlibrium in this case is achieved when
the total chemical potential $\mu = \tilde{\mu}(P,T) + \mathcal{V}$ is constant inside the system (in our case, inside NS; see below). The 
results of Ref.~\cite[c.~73--74]{LanLif02} can be straightforwardly generalized to the case of a system consisting of two types of particles:
left and right electrons. In this situation we obtain that total chemical potentials should coincide in the state of equilibrium:
$\mu_{\mathrm{L}} = \mu_{\mathrm{R}}$.

The chemical potential is defined as the energy acquired by a system when one particle is added there~\cite[p.~71]{LanLif02}.
In the present work, NS serves as a system. Thus the chemical potential should be defined with respect to vacuum, which is the space outside NS,
where there is no background matter. The quantities $\mu_{\mathrm{L,R}}$, used in the present work, have the meaning of the total chemical
potentials including the interaction with matter, which is the analogue of the external field. It can be seen in Eqs.~\eqref{eq:Wew} and~\eqref{eq:nmuV}
since the energies of left and right electrons in the distribution functions are defined with respect to vacuum.

Moreover the formal redefinition of the chemical potential, proposed in Ref.~\cite{SigLei16}:
${\mu}_{\mathrm{L,R}} \to \tilde{\mu}_{\mathrm{L,R}} = {\mu}_{\mathrm{L,R}} - {V}_{\mathrm{L,R}}$, which would be meaningful only inside NS, is
unlikely to be implemented in practice. This redefinition is equivalent to the independent choice of the zero energy for left and right electrons.
However, if $\Gamma_f \neq 0$, left and right particles collide with protons and arrive to the state of a thermodynamic equilibrium.
Thus left and right particles do not form two independent thermodynamic systems. Hence it is impossible to shift simultaneously chemical
potential by two different values $V_{\mathrm{L}} \neq V_{\mathrm{R}}$.

Note that, in vacuum, there is no energy splitting of relativistic particles with opposite helicities; cf. Eq.~\eqref{eq:enlev}.
Hence, outside NS, one can choose equal zero energy levels for left and right particles. Thus, using total chemical potentials, including the interaction
with matter and defined with respect to vacuum, is preferred.

Therefore Eq.~\eqref{eq:siglwrong} proposed in Ref.~\cite{SigLei16} neither is  confirmed by the direct calculation of the probability of the
processes $e_\mathrm{L,R} \leftrightarrow e_\mathrm{R,L}$ in Sec.~\ref{sub:MAT}, nor is in agreement with the results of the macroscopic
thermodynamics.

\section{Generation of magnetic fields in magnetars\label{sec:EVOL}}

In this section we shall numerically solve Eqs.~(\ref{eq:heq})-(\ref{eq:mu5eq})
accounting for the new dependence of $\Gamma_{f}$ on $T$ in
Eq.~(\ref{eq:m5kincorr}). Previously analogous problem was studied in
Refs.~\cite{DvoSem15b,DvoSem15c}. It is necessary to briefly recall the initial condition for Eqs.~(\ref{eq:heq})-(\ref{eq:mu5eq}).

We shall adopt the initial Kolmogorov spectrum of the magnetic energy
density $\rho_{\mathrm{B}}(k,t_{0})=\mathcal{C}k^{-5/3}$, where the
constant $\mathcal{C}$ is related by Eq.~(\ref{eq:hdef}) to the
seed field $B(t_{0})=B_{0}=10^{12}\thinspace\text{G}$ typical in
a young pulsar. The integration in Eq.~(\ref{eq:hdef}) is in the
range $k_{\mathrm{min}}<k<k_{\mathrm{max}}$, where $k_{\mathrm{min}}=2\times10^{-11}\thinspace\text{eV}=R_{\mathrm{NS}}^{-1}$,
$R_{\mathrm{NS}}=10\thinspace\text{km}$ is the NS radius, $k_{\mathrm{max}}=\Lambda_{\mathrm{B}}^{-1}$,
and $\Lambda_{\mathrm{B}}$ is the minimal scale of the magnetic field
generated, which is a free parameter. The initial spectrum of the
magnetic helicity density is taken in the form, $h(k,t_{0})=2q\rho_{\mathrm{B}}(k,t_{0})/k$,
where $0\leq q\leq1$ is the parameter defining the initial helicity:
$q=0$ corresponds to the initially nonhelical field and $q=1$ to
the field with the maximal helicity. We shall choose the initial value of the chiral
imbalance in the following way: $\mu_{5}(t_{0})=1\thinspace\text{MeV}$. Note that the evolution of the magnetic field is almost insensitive to
$\mu_{5}(t_{0})$ because of the huge $\Gamma_f$.

The number densities of electrons, protons and neutrons will be taken
as $n_{e}=n_{p}=9\times10^{36}\thinspace\text{cm}^{-3}$ and $n_{n}=1.8\times10^{38}\thinspace\text{cm}^{-3}$.
It corresponds to $\mu_{e}=125\thinspace\text{MeV}$ since electrons are
ultrarelativistic in NS. These particle densities can be found in
a typical NS.

To account for the energy balance in the system consisting of the magnetic field and the background matter, one should quench the
parameter $\Pi$ in Eq.~(\ref{eq:JCS})
(see Appendix~\ref{sec:ENSOR}),
\begin{equation}\label{eq:Piq}
  \Pi \to \Pi
  \left[
    1-\frac{B^{2}}{B_{\mathrm{eq}}^{2}(T)}
  \right],
\end{equation}
where $B$ and $B_{\mathrm{eq}}$ are given in Eqs.~(\ref{eq:hdef}) and~\eqref{Beq}. The quenching in Eq.~(\ref{eq:Piq})
allows one to prevent the excessive growth of the
magnetic field at $t\gg t_0$. At $B \ll B_{\mathrm{eq}}$, the quenching in Eq.~(\ref{eq:Piq}) is equivalent to that in Ref.~\cite{DvoSem15c}.

We shall study the evolution of the magnetic field in a thermally relaxed NS at $t_{0}<t\lesssim10^{6}\thinspace\text{yr}$,
where $t_{0}\sim10^{2}\thinspace\text{yr}$. If one studies NS with a rather small mass $M < 1.44 M_\odot$, where $M_\odot = 2 \times 10^{33}\,\text{g}$
is the solar mass, then, as shown in Ref.~\cite{GneYakPot01}, in the time interval, NS cools down owing to the neutrino emission in modified
Urca-processes. It results in the dependence of the temperature on time~\cite{GneYakPot01,Pet92},
\begin{equation}\label{eq:cooling}
  T(t)=T_{0}
  \left(
    \frac{t}{t_{0}}
  \right)^{-1/6},
\end{equation}
where $T_{0}=(10^{8}-10^{9})\thinspace\text{K}$ is the temperature at $t=t_{0}$. For more massive NSs, the cooling due to the neutrino emission
can become faster than it results from Eq.~\eqref{eq:cooling}. Moreover, as found in Ref.~\cite{GneYakPot01}, for NS with the mass
$M = 1.3 M_\odot$ at $t_0 = 10^2\,\text{yr}$, the temperature in the center of NS is $T_0 = 4\times 10^8\,\text{K}$ at the absence
of the superfluidity of the neutron component. The central temperature is $T_0 = 3\times 10^8\,\text{K}$, if the superfluidity
is present only in the crust of NS. In the situation when there is a superfluidity in the core of NS, $T_0$ can be significantly less than $10^8\,\text{K}$.
Therefore we shall consider rather light NS either totally without a superfluidity or when only the crust is superfluid.
Using the results of Ref.~\cite{Kel73}, one gets the time dependence
of the conductivity as $\sigma_{\mathrm{cond}}(t)=\sigma_{0}\left(t/t_{0}\right)^{1/3}$,
where $\sigma_{0}=2.7\times10^{8} \times (T_0/10\,\text{K})^{-2} \thinspace\text{MeV}$ is the conductivity at 
$t=t_{0}$. We shall also take into account the corrected temperature dependence of $\Gamma_{f}$ in Eq.~(\ref{eq:m5kincorr}),
\begin{equation}\label{eq:Gfcooling}
  \Gamma_{f} = 1.6\times10^{11}\thinspace\text{s}^{-1}
  \left(
    \frac{t}{t_{0}}
  \right)^{-1/6},
\end{equation}
where we account for Eq.~(\ref{eq:cooling}) and the chosen value of the electron number density.

\begin{figure}
  \centering
  \subfigure[]
  {\label{2a}
  \includegraphics[scale=.23]{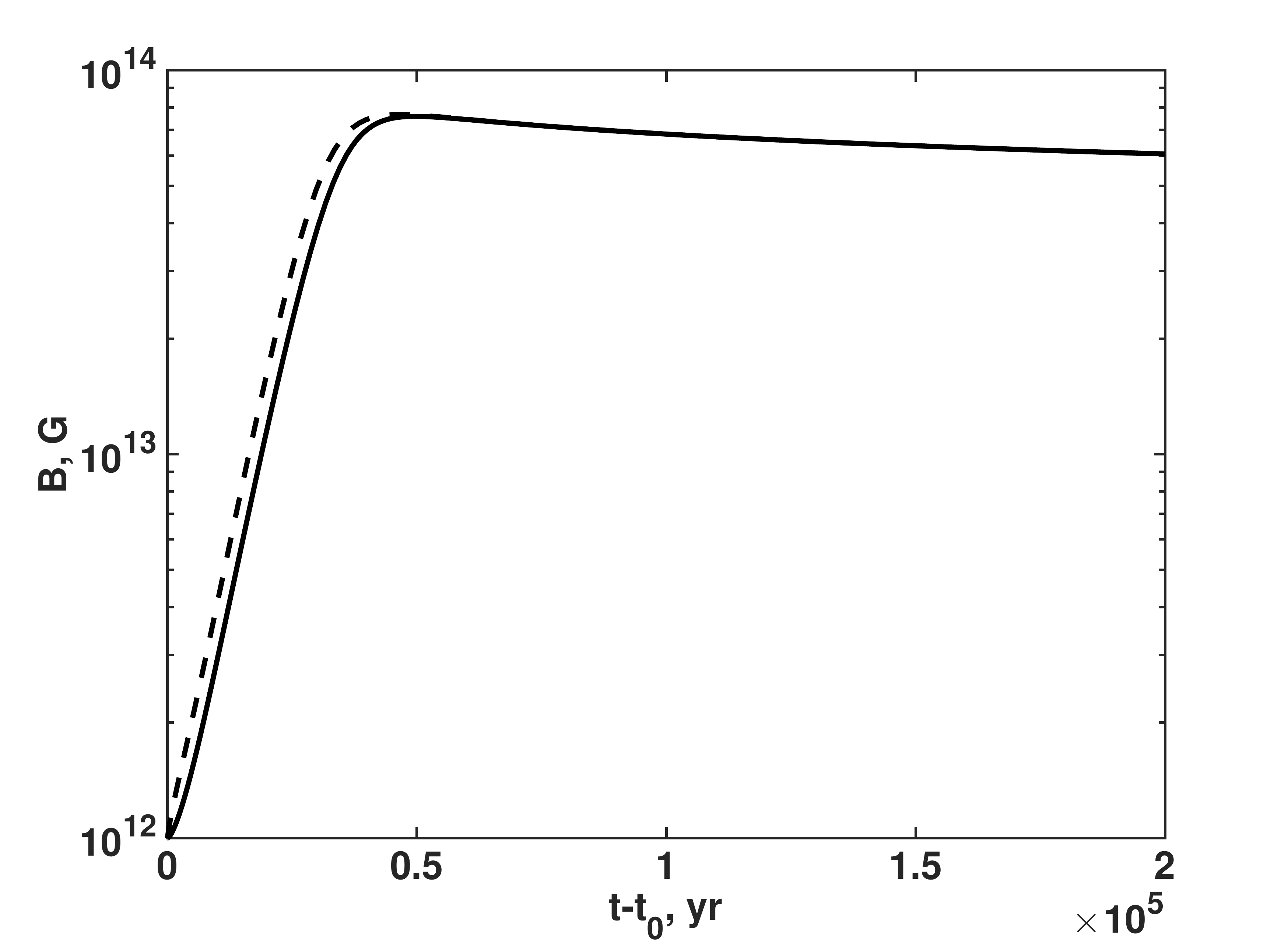}}
  \hskip-.7cm
  \subfigure[]
  {\label{2b}
  \includegraphics[scale=.23]{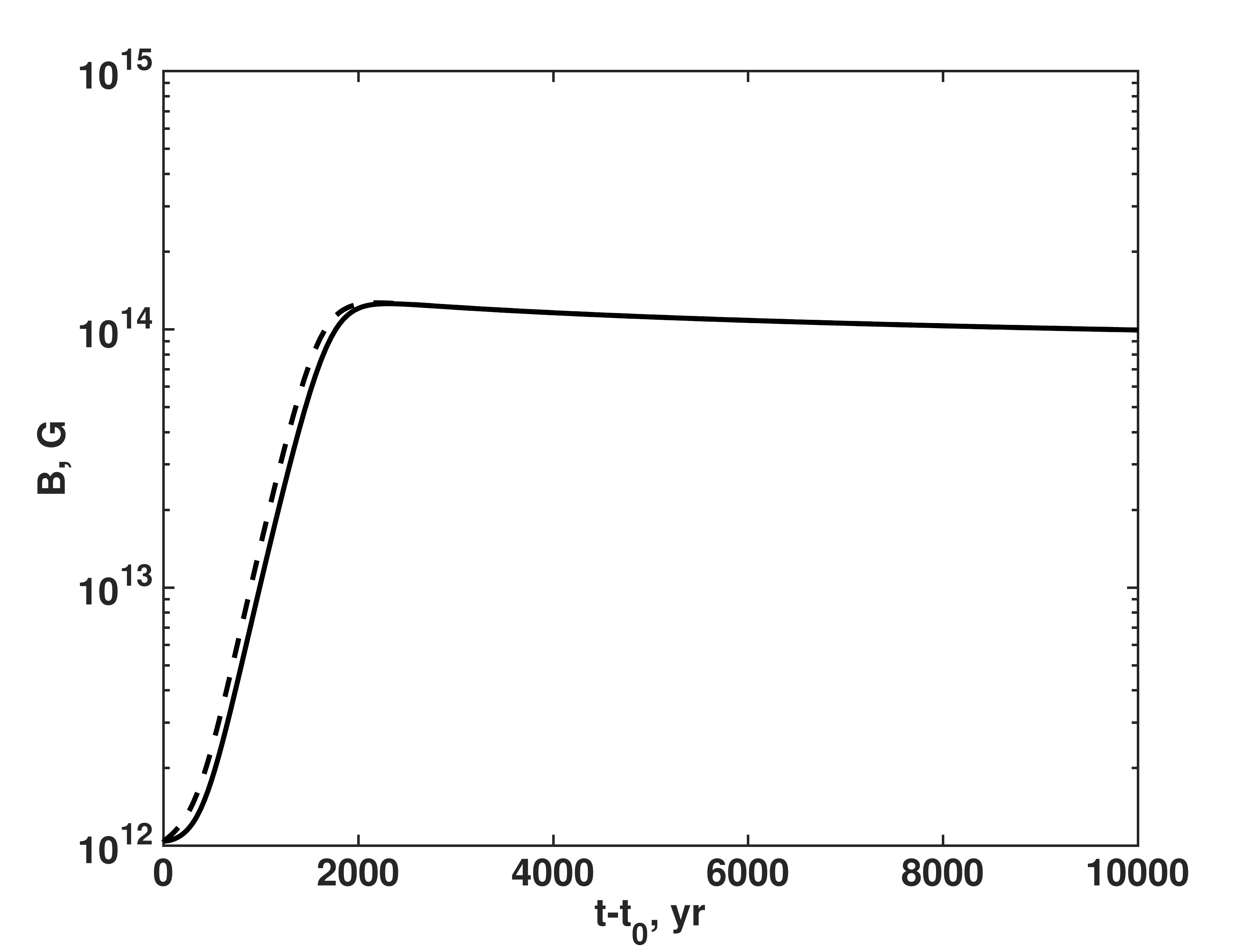}}
  \\
  \subfigure[]
  {\label{2c}
  \includegraphics[scale=.23]{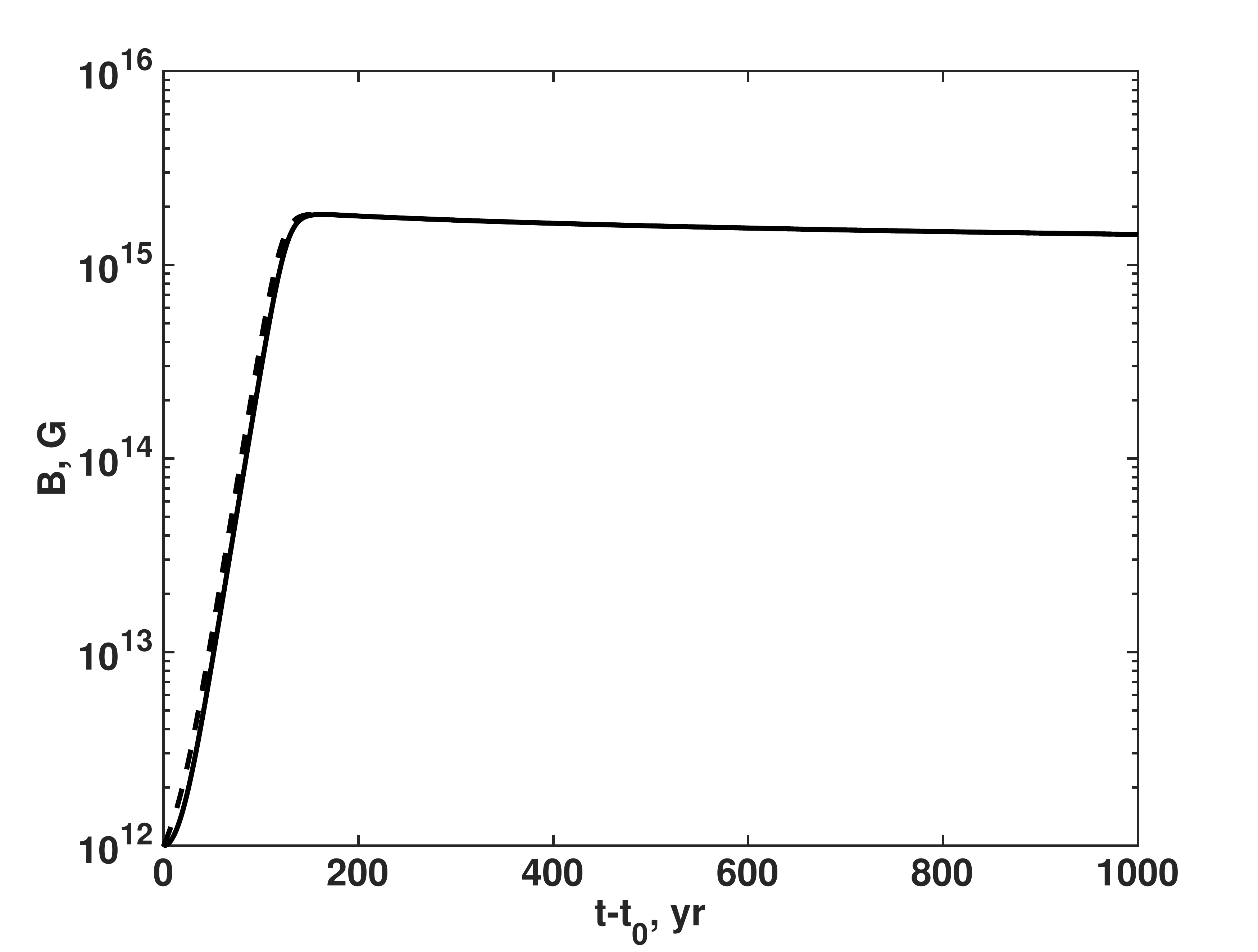}}
  \hskip-.7cm
  \subfigure[]
  {\label{2d}
  \includegraphics[scale=.23]{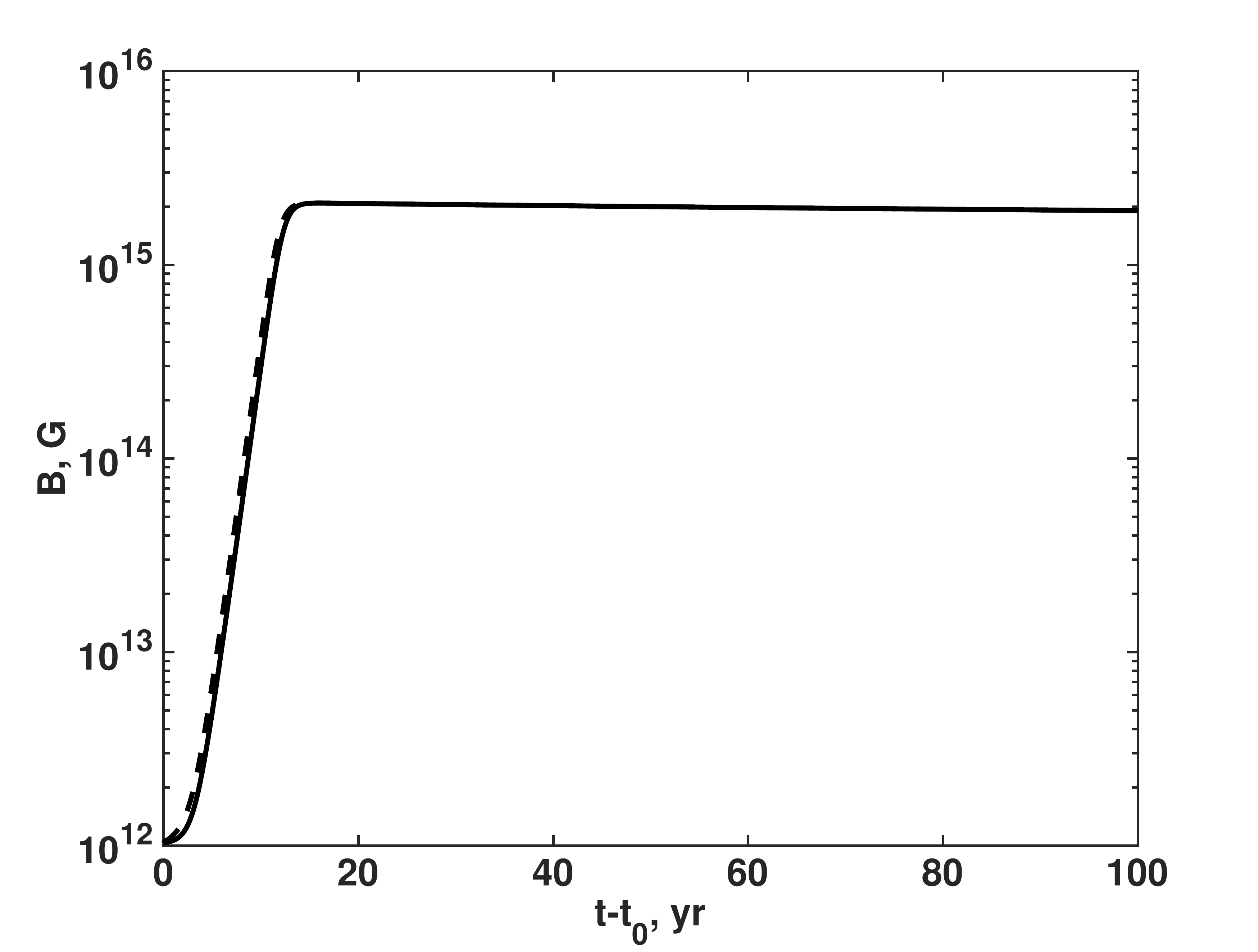}}
  \protect\caption{
  The evolution of magnetic fields in magnetars versus $t-t_{0}$
  for different $k_{\mathrm{max}}$ and $T_0$.
  Solid lines correspond to initially nonhelical
  fields with $q=0$ and dashed lines to maximally helical fields with
  $q=1$.
  (a)~The magnetic field evolution for  
  $k_{\mathrm{max}}=2\times10^{-10}\thinspace\text{eV}$
  ($\Lambda_{\mathrm{B}}=1\thinspace\text{km}$) and
  $T_0 = 10^8\,\text{K}$.
  (b)~The behavior of the magnetic field with   
  $k_{\mathrm{max}}=2\times10^{-9}\thinspace\text{eV}$
  ($\Lambda_{\mathrm{B}}=10^2\thinspace\text{m}$) and
  $T_0 = 10^8\,\text{K}$.
  (c)~The magnetic field evolution for  
  $k_{\mathrm{max}}=2\times10^{-10}\thinspace\text{eV}$
  ($\Lambda_{\mathrm{B}}=1\thinspace\text{km}$) and
  $T_0 = 10^9\,\text{K}$.
  (d)~The behavior of the magnetic field with   
  $k_{\mathrm{max}}=2\times10^{-9}\thinspace\text{eV}$
  ($\Lambda_{\mathrm{B}}=10^2\thinspace\text{m}$) and
  $T_0 = 10^9\,\text{K}$.  
  \label{fig:B}}
\end{figure}

In Fig.~\ref{fig:B} we show the time dependence of the magnetic field on the basis of the numerical solution of Eqs.~(\ref{eq:heq})-(\ref{eq:mu5eq})
with the chosen initial conditions. One can see in Fig.~\ref{fig:B} that the magnetic field grows exponentially at small evolution times.
This growth is driven by the nonzero $V_5$ induced by the electroweak $eN$ interaction.

Magnetic fields grow at $t-t_0 \sim (10-10^5)\,\text{yr}$ depending on $\Lambda_\mathrm{B}$ and $T_0$.
The fastest growth takes place at $T_0 = 10^9\,\text{K}$, which corresponds to the smallest $\sigma_{\mathrm{cond}}$, as well as for small scale magnetic fields. This fact can be explained
basing on the Faraday equation,
\begin{equation}\label{FE}
  \frac{\partial\mathbf{B}}{\partial t} =
  \frac{\Pi}{\sigma_{\mathrm{cond}}}
  \left(
    \nabla\times\mathbf{B}
  \right) +
  \frac{1}{\sigma_{\mathrm{cond}}}\nabla^{2}\mathbf{B},
\end{equation}
which is equivalent to Eqs.~\eqref{eq:heq} and~\eqref{eq:rhoeq}. As follows from Eq.~\eqref{FE}, the typical magnetic field growth time is
$t \sim \sigma_{\mathrm{cond}} \Lambda_\mathrm{B}/\Pi$, which explains the aforementioned feature. Note that the time for the magnetic
field to reach the maximal strength is $t \sim (10^3 - 10^5)\,\text{yr}$ at $T_0 = 10^8\,\text{K}$, see Figs.~\ref{2a} and~\ref{2b},
which is close to the observed ages of young magnetars~\cite{MerPonMel15}.

The maximal magnetic field strength $B_\mathrm{max} \sim (10^{14}-10^{15}) \,\text{G}$ is characterized by the initial thermal energy of background fermions.
After reaching $B_\mathrm{max}$, magnetic fields start to decrease slowly. It results from the continuous energy loss of NS by the neutrino emission.
Note, that $B_\mathrm{max}$ in Fig.~\ref{2a} is less than that in Fig.~\ref{2b}. This feature is a consequence of the greater time scale in Fig.~\ref{2a}.
Thus neutrinos will carry away more energy from NS. It should be also mentioned that $B_\mathrm{max} \sim 10^{15} \,\text{G}$
in Figs.~\ref{2c} and~\ref{2d} is close to the magnetic field strength predicted in magnetars~\cite{MerPonMel15}.

In Fig.~\ref{fig:B}, we depict the evolution of magnetic fields with different initial helicities.
One can see that the discrepancy in the behavior of such fields is noticeable only at small evolution times.
This result is in agreement with the findings of Ref.~\cite{DvoSem15b}.

Note that the new dependence of $\Gamma_{f}$ on time in Eq.~\eqref{eq:Gfcooling}
does not significantly influence the evolution of magnetic fields compared to the results of Refs.~\cite{DvoSem15b,DvoSem15c}.
As found in Ref.~\cite{DvoSem15a}, almost any initial $\mu_{5}(t_{0})$
is washed out very rapidly owing to the great value of $\Gamma_{f}$.
Hence, despite $\Gamma_{f}(t_{0})$ is different from that used
in Refs.~\cite{DvoSem15b,DvoSem15c}, we cannot expect a sizable
discrepancy in the behavior of magnetic fields at $t\gtrsim t_{0}$.
At greater evolution times $t\gg t_0$, $\Gamma_{f}(t)$ will decrease slower than
that in Refs.~\cite{DvoSem15b,DvoSem15c}. However, in this time interval, the evolution of magnetic
fields is affected mainly by the quenching of $\Pi$ in Eq.~(\ref{eq:Piq}).

\section{Conclusion\label{sec:CONCL}}

In conclusion we mention that in this paper we have studied $ep$
collisions in dense matter of NS. In particular, in
Sec.~\ref{sec:COLLISIONS}, we have considered the scattering
of polarized electrons off unpolarized protons. In Sec.~\ref{sub:MAT}, we have computed
the total helicity flip rate of an electron when it collides with a proton,
where we have exactly accounted for the electroweak interaction between electrons
and neutrons in the NS matter. The kinetic equation for the chiral imbalance has been
derived in Sec.~\ref{sub:KIN}. The relaxation of the chiral imbalance from the point of view
of thermodynamics has been analyzed in Sec.~\ref{sub:TERM}. The obtained results have been used
in Sec.~\ref{sec:EVOL} for the description of the magnetic fields generation in magnetars.

Note that initially the model for the generation of the magnetic field in magnetars driven by the 
electroweak $eN$ interaction was proposed in Ref.~\cite{DvoSem15a}. Then, in Refs.~\cite{DvoSem15b,DvoSem15c}, this model
was corrected. However the helicity flip rate in $ep$ collisions was estimated in Refs.~\cite{DvoSem15a,DvoSem15b} basing
on qualitative ideas of classical physics~\cite[pp.~66--67]{AleBogRuk84}. The particle spin is known to be
a purely quantum object. That is why its evolution
should be treated appropriately. In the present work we used the QFT methods to compute the helicity flip rate. It can explain
the discrepancy of our results from those in Refs.~\cite{DvoSem15a,DvoSem15b};
cf. Eqs.~(\ref{eq:m5kincorr}) and~(\ref{eq:Gfsimp}).
  
Another important result obtained in the present work was the analysis
of the influence of the electroweak interaction of electrons
with background nucleons on the helicity flip process in $ep$ collisions. Using the
method of the exact solutions of the Dirac equation in an external field (see Appendix~\ref{sec:SOLDIREQ}),
assuming that the scattering is elastic, as well as supposing that electrons are ultrarelativistic, we have found
that the effective potentials $V_\mathrm{L,R}$ do not contribute explicitly to the expression of the total
probability of the processes $e_\mathrm{L,R} \leftrightarrow e_\mathrm{R,L}$ in Eq.~\eqref{eq:Wfin}.
Hence the kinetic equation for the chiral imbalance in Eq.~\eqref{eq:m5kincorr} coincides with
that in Refs.~\cite{DvoSem15a,DvoSem15b}, contrary to the recent claim in Ref.~\cite{SigLei16}; cf. Eq.~\eqref{eq:siglwrong}.
Moreover our kinetic Eq.~\eqref{eq:m5kincorr} is confirmed by the laws of thermodynamics; cf. Sec.~\ref{sub:TERM}. Therefore, it
is the electroweak $eN$ interaction which drives the magnetic field growth in the model in Refs.~\cite{DvoSem15a,DvoSem15b,DvoSem15c}.

Finally, the accurate account for the energy conservation made in Appendix~\ref{sec:ENSOR} allowed one to modify the quenching of the
parameter $\Pi$ in Eq.~\eqref{eq:Piq} compared to the results of Ref.~\cite{DvoSem15c}. This fact led to a more adequate description
of the magnetic field evolution in magnetars in Sec.~\ref{sec:EVOL}, especially at $B \sim B_\mathrm{eq}$. Nevertheless
the strengths of the magnetic field generated $B_\mathrm{max} \sim (10^{14}-10^{15})\,\text{G}$ and the time of the magnetic
field growth $\lesssim 5 \times 10^{4}\,\text{yr}$
are in agreement with astrophysics predictions for magnetars~\cite{MerPonMel15}.

It is interesting to compare the results of the present work with the recent findings in Ref.~\cite{Yam15}, where the
superstrong magnetic fields with $B\sim 10^{18}\,\text{G}$ are generated in magnetars owing to the combination of the chiral magnetic
and chiral vortical effects~\cite{KhaLiaVol16} in the electron-neutrino medium during $\sim 10^{-23}\,\text{s}$.
Note that the magnetic field length scale, predicted in Ref.~\cite{Yam15}, is $\sim 10^{-12}\,\text{cm}$. It is claimed in Ref.~\cite{Yam15} that
the length scale can increase due to a mechanism analogous to the inverse cascade of energy~\cite{DanGur00}. However no
quantitative estimates of the length scale enhancement are provided in Ref.~\cite{Yam15}. Moreover, in Ref.~\cite{MoiBis08} it was shown that
small scale magnetic fields dissipate effectively in a time interval of several seconds because of the magnetic field reconnection.
Therefore the application of the results of Ref.~\cite{Yam15} for the explanation of magnetic fields in magnetars looks questionable.
The same argument can be put forward with respect to the results of Ref.~\cite{SigLei16}, since in that work, the generation
of small scale magnetic fields was predicted. As follows from the results of the numerical simulation in Sec.~\ref{sec:EVOL},
magnetic fields predicted in our model are large scale, that makes them insensitive to dissipation processes like the magnetic reconnection.
 
I am thankful to V.~G.~Bagrov and V.~B.~Semikoz for fruitful discussions. This work was supported by RFBR (grant No.~15-02-00293), DAAD (grant No.~91610946)
and the Tomsk State University Competitiveness Improvement Program.

\appendix

\section{Solution of the Dirac equation for an electron, electroweakly interacting
with nuclear matter\label{sec:SOLDIREQ}}

In this Appendix we present the exact solution of the Dirac equation
for an electron electroweakly interacting with nuclear matter consisting
of neutrons and protons. Note that previously this problem was studied
in Ref.~\cite{Gri08}. Nevertheless here we present this solution
in a way convenient for subsequent computations.

Let us consider the electroneutral matter in NS consisting of neutrons,
protons, and electrons. This matter is supposed to be at rest and
unpolarized. The Dirac equation for a test electron, described by
the bispinor wave function $\psi$, electroweakly interacting with
neutrons and protons, has the form,
\begin{equation}\label{eq:Direq}
  \left[
    \mathrm{i}\gamma^{\mu}\partial_{\mu} -m-\gamma^{0}
    \left(
      V_{\mathrm{L}}P_{\mathrm{L}}+V_{\mathrm{R}}P_{\mathrm{R}}
    \right)
  \right]
  \psi=0,
\end{equation}
where
\begin{align}\label{eq:VLR}
  V_{\mathrm{L}} = & \frac{G_{\mathrm{F}}}{\sqrt{2}}
  \left[
    n_{n}-n_{p}(1-4\xi)
  \right](1-2\xi),
  \nonumber
  \\
  V_{\mathrm{R}}= & -\frac{G_{\mathrm{F}}}{\sqrt{2}}
  \left[
    n_{n}-n_{p}(1-4\xi)
  \right]
  2\xi,
\end{align}
are the effective potentials of the interaction of left and right
chiral projections with matter, $n_{n,p}$ are the constant and uniform densities
of neutrons and protons, $\xi=\sin^{2}\theta_{\mathrm{W}}\approx0.23$
is the Weinberg parameter, $P_{\mathrm{L,R}}=(1\mp\gamma^{5})/2$
are the chiral projectors, and $\gamma^{5} = \mathrm{i}\gamma^{0}\gamma^{1}\gamma^{2}\gamma^{3}$.

We shall look for the solution of Eq.~(\ref{eq:Direq}) in the form,
\begin{equation}
  \psi=\frac{1}{\sqrt{V}}\exp[-\mathrm{i}Et+\mathrm{i}(\mathbf{pr})]u,
\end{equation}
where $u$ is the constant spinor. Then, using Eq.~(\ref{eq:Direq}),
we obtain the energy spectrum in the form,
\begin{equation}\label{eq:enlev}
  E = \bar{V}+E_{0},
  \quad
  \bar{V} = \frac{V_{\mathrm{L}}+V_{\mathrm{R}}}{2},
  \quad
  E_{0}^{2} =
  \left(
    p-sV_{5}
  \right)^{2}
  + m^{2},
  \quad
  V_{5}=\frac{V_{\mathrm{L}}-V_{\mathrm{R}}}{2},
\end{equation}
where $s=+1$ for right electron and $s=-1$ for left ones. Note that in~\eqref{eq:enlev} we do not take into account the positron degrees of freedom. If electrons
are ultrarelativistic, one gets from Eq.~(\ref{eq:enlev}) that $E_{\pm}=p+V_{\mathrm{R},\mathrm{L}}$.

Using the chiral representation for
the Dirac matrices~\cite[pp.~691--696]{ItzZub80}, one can find the basis bispinors in the form,
\begin{equation}\label{eq:upm}
  u_{+} = N_{+}
  \left(
    \begin{array}{c}
      w_{+} \\
      -\frac{m}{E_{+}+p-V_{\mathrm{L}}}w_{+}
    \end{array}
  \right),
  \quad
  u_{-} = N_{-}
  \left(
    \begin{array}{c}
      -\frac{m}{E_{-}+p-V_{\mathrm{R}}}w_{-}
      \\
      w_{-}
    \end{array}
  \right),
\end{equation}
where $w_{\pm}$ are the helicity amplitudes~\cite[p.~86]{BerLifPit82},
\begin{equation}\label{eq:helampl}
  w_{+}(\mathbf{p}) =
  \left(
    \begin{array}{c}
      e^{-\mathrm{i}\phi/2}\cos\frac{\vartheta}{2}
      \\
      e^{\mathrm{i}\phi/2}\sin\frac{\vartheta}{2}
    \end{array}
  \right)
  \quad
  w_{-}(\mathbf{p}) =
  \left(
    \begin{array}{c}
      -e^{-\mathrm{i}\phi/2}\sin\frac{\vartheta}{2}
      \\
      e^{\mathrm{i}\phi/2}\cos\frac{\vartheta}{2}
    \end{array}
  \right).
\end{equation}
Here $\phi$ and $\vartheta$ are the spherical angles giving the direction
of the vector $\mathbf{p}$. The spinors $w_{\pm}$ are the eigenvectors
of the helicity operator: $\left( \bm{\sigma}\cdot\mathbf{p} \right) w_{\pm} = \pm |\mathbf{p}| w_{\pm}$.

If we normalize the electron wave function in the following way:
\begin{equation}\label{eq:norm}
  \int\mathrm{d}^{3}x\psi^{\dagger}\psi=1,
\end{equation}
we can find that the normalization constants $N_{\pm}$ in Eq.~(\ref{eq:upm})
are equal to
\begin{equation}\label{eq:normcoef}
  N_{\pm} = \sqrt{\frac{E_{\pm}+p-V_{\mathrm{L},\mathrm{R}}}{2E_{0\pm}}},
\end{equation}
where $E_{\pm}$ and $E_{0\pm}$ are given in Eq.~(\ref{eq:enlev}).

\section{Calculation of integrals over the phase space\label{sec:INTCALC}}

In this Appendix we compute the integrals over the momenta of electrons and protons in Eq.~\eqref{eq:Wew}. The computation of integrals in Eq.~\eqref{eq:Wew} can be made independently since the matrix element in Eq.~\eqref{eq:M2ew} is factorized in the approximation of elastic scattering.

First, let us compute the integral over the electron momenta,
\begin{align}\label{Ie}
  I_{e}= & \int
  \frac{\mathrm{d}^{3}p_{1}\mathrm{d}^{3}p_{2}
  \left(
    p_{1}+p_{2}
  \right)^{2}
  \left[
    1-
    \left(
      \mathbf{n}_{1}\cdot\mathbf{n}_{2}
    \right)
  \right]}{
  16
  \left(
    p_{1}-V_{5}
  \right)^{2}
  \left(
    p_{2}+V_{5}
  \right)^{2}}
  \delta^{3}
  \left(
    \mathbf{p}_{1}-\mathbf{p}_{2}-\mathbf{q}
  \right)
  \nonumber
  \\
  & \times
  \delta
  \left(
    p_{1} - p_{2} + \mathcal{E}_1 - \mathcal{E}_2 +
    V_{\mathrm{R}}  - V_{\mathrm{L}}
  \right)
  \theta
  \left(
    \mu_{\mathrm{R}}-p_{1}-V_{\mathrm{R}}
  \right)
  \theta
  \left(
    p_{2}+V_{\mathrm{L}}-\mu_{\mathrm{L}}
  \right),
\end{align}
where $\mathbf{q} = \mathbf{k}_2 - \mathbf{k}_1$. In Eq.~\eqref{Ie}
we assume that electrons are highly degenerate and ultrarelativistic. Using the conservation laws, one can rewrite Eq.~\eqref{Ie} in the form,
\begin{equation}\label{eq:Ieew}
  I_{e} = 
  \pi\frac{|\mathbf{q}|^{2}-V_{5}^{2}}{4|\mathbf{q}|}
  \frac{
  \left(
    \mu_{\mathrm{R}}-\mu_{\mathrm{L}}
  \right)
  \theta
  \left(
    \mu_\mathrm{R}-\mu_\mathrm{L}
  \right)}{
  \left(
    \mu_{\mathrm{R}}-\bar{V}
  \right)
  \left(
    \mu_{\mathrm{L}}-\bar{V}
  \right)}
  \approx
  \frac{\pi|\mathbf{q}|}{4\mu_{e}^{2}}
  \left(
    \mu_{\mathrm{R}}-\mu_{\mathrm{L}}
  \right)
  \theta
  \left(
    \mu_\mathrm{R}-\mu_\mathrm{L}
  \right),
\end{equation}
where we suppose that $\mu_{\mathrm{L,R}}\gg\bar{V}$, $\mu_{\mathrm{L}}\approx\mu_{\mathrm{R}}\approx\mu_{e}$,
and $|\mathbf{q}|\gg V_{5}$. The important consequence of Eq.~\eqref{eq:Ieew}, is the fact that $V_{5}$ does not contribute to the difference of the chemical potentials in the numerator.

The integral over the proton momenta,
\begin{equation}\label{eq:IpqQ}
  I_{p}= \int
  \frac{\mathrm{d}^{3}k_{1}\mathrm{d}^{3}k_{2}}
  {\mathcal{E}_{1}\mathcal{E}_{2}}
  \frac{\mathcal{E}_{1}\mathcal{E}_{2}+M^{2} +
  \left(
    \mathbf{k}_{1}\cdot\mathbf{k}_{2}
  \right)}  
  {
  \left[
    \left(
      \mathbf{k}_{1}-\mathbf{k}_{2}
    \right)^{2} +
    \omega_{p}^{2}
  \right]^{3/2}}
  f_{p}(\mathcal{E}_{1}-\mu_{p})
  \left[
    1-f_{p}(\mathcal{E}_{2}-\mu_{p})
  \right],
\end{equation}
can be computed by changing the integration variables: $\mathbf{q}=\mathbf{k}_{2}-\mathbf{k}_{1}$ and $\mathbf{Q}=\mathbf{k}_{2}+\mathbf{k}_{1}$.
Note that we insert the plasma frequency in a degenerate matter~\cite{BraSeg93}  $\omega_{p}^{2}=4\alpha_{\mathrm{em}}\mu_{e}^{2}/3\pi$ to Eq.~\eqref{eq:IpqQ} to avoid the infrared divergence. Supposing that protons are degenerate and have a low but nonzero temperature, as well as the NS matter is electroneutral, we can transform Eq.~\eqref{eq:IpqQ} to the form,
\begin{equation}\label{eq:IpKE}
  I_{p} = 16 \mu_{e} M \pi^{2}T
  \left[
    \ln
    \left(
      \frac{48\pi}{\alpha_{\mathrm{em}}}
    \right)
    - 4
  \right].
\end{equation}
Using Eqs.~\eqref{eq:Ieew} and~\eqref{eq:IpKE}, we obtain Eq.~\eqref{eq:Wfin}

\section{Kinetic equations for the occupation numbers of left and right electrons\label{sec:KINEQ}}

In this Appendix, basing on the Boltzmann equation with the collision integral, we derive the kinetic Eq.~\eqref{eq:kineqN}.

At the absence of external fields, the kinetic equations for the spatially homogeneous distribution functions of left and right electrons $f_\mathrm{L,R} = f_\mathrm{L,R}(\mathbf{p}_\mathrm{L,R},t)$ have the form,
\begin{equation}\label{eq:kineqLR}
  \frac{\partial f_\mathrm{L,R}}{\partial t} =
  J_\mathrm{coll}
  \left[
    f_\mathrm{L,R}
  \right],
\end{equation}
where $J_\mathrm{coll}\left[f_\mathrm{L,R}\right]$ are the collision integrals.
Since, in Secs.~\ref{sub:MAT} and~\ref{sub:KIN}, we discuss collisions in which the helicity of electrons flips, then $J_\mathrm{coll}\left[f_\mathrm{L,R}\right]$ can be represented in the following way~\cite{GroLeeWee83}:
\begin{align}\label{eq:StLR}
  J_\mathrm{coll}
  \left[
    f_\mathrm{L}
  \right]
  = &
  \int\frac{\mathrm{d}^3p_\mathrm{R}}{(2\pi)^{3}}
  \frac{\mathrm{d}^3 k_1}{(2\pi)^{3}}
  \frac{\mathrm{d}^3 k_2}{(2\pi)^{3}}
  \frac{|\mathcal{M}_{R\to L}|^{2}}{2\mathcal{E}_1\mathcal{E}_2}
  (2\pi)^4\delta^{4}
  \left(
    p_\mathrm{L}+k_2-p_\mathrm{R}-k_1
  \right)
  \notag
  \\
  & \times
  \left(
    1-f_\mathrm{L}
  \right)
  f_\mathrm{R}
  \left[
    1-f_{p}(\mathbf{k}_2)
  \right]
  f_{p}(\mathbf{k}_1)
  \nonumber
  \\
  & -
  \int\frac{\mathrm{d}^3p_\mathrm{R}}{(2\pi)^{3}}
  \frac{\mathrm{d}^3 k_1}{(2\pi)^{3}}
  \frac{\mathrm{d}^3 k_2}{(2\pi)^{3}}
  \frac{|\mathcal{M}_{L\to R}|^{2}}{2\mathcal{E}_1\mathcal{E}_2}
  (2\pi)^4\delta^{4}
  \left(
    p_\mathrm{L}+k_1-p_\mathrm{R}-k_2
  \right)
  \notag
  \\
  & \times
  f_\mathrm{L}
  \left(
    1-f_\mathrm{R}
  \right)
  \left[
    1-f_{p}(\mathbf{k}_2)
  \right]
  f_{p}(\mathbf{k}_1),
  \nonumber
  \\
  J_\mathrm{coll}
  \left[
    f_\mathrm{R}
  \right]
  = &
  \int\frac{\mathrm{d}^3p_\mathrm{L}}{(2\pi)^{3}}
  \frac{\mathrm{d}^3 k_1}{(2\pi)^{3}}
  \frac{\mathrm{d}^3 k_2}{(2\pi)^{3}}
  \frac{|\mathcal{M}_{L\to R}|^{2}}{2\mathcal{E}_1\mathcal{E}_2}
  (2\pi)^4\delta^{4}
  \left(
    p_\mathrm{L}+k_1-p_\mathrm{R}-k_2
  \right)
  \notag
  \\
  & \times
  \left(
    1-f_\mathrm{R}
  \right)
  f_\mathrm{L}
  \left[
    1-f_{p}(\mathbf{k}_2)
  \right]
  f_{p}(\mathbf{k}_1)
  \nonumber
  \\
  & -
  \int\frac{\mathrm{d}^3p_\mathrm{L}}{(2\pi)^{3}}
  \frac{\mathrm{d}^3 k_1}{(2\pi)^{3}}
  \frac{\mathrm{d}^3 k_2}{(2\pi)^{3}}
  \frac{|\mathcal{M}_{R\to L}|^{2}}{2\mathcal{E}_1\mathcal{E}_2}
  (2\pi)^4\delta^{4}
  \left(
    p_\mathrm{L}+k_2-p_\mathrm{R}-k_1
  \right)
  \notag
  \\
  & \times
  f_\mathrm{R}
  \left(
    1-f_\mathrm{L}
  \right)
  \left[
    1-f_{p}(\mathbf{k}_2)
  \right]
  f_{p}(\mathbf{k}_1),
\end{align}
where $f_{p}(\mathbf{k}_{1,2})$ are distribution functions of incoming and outgoing protons, which are defined in Sec.~\ref{sub:MAT}, $\mathcal{M}_{L\to R}$ and $\mathcal{M}_{R\to L}$ are the matrix elements of the corresponding processes. The normalization of these quantities coincides with that in Eq.~\eqref{eq:matrel}. We account for the averaging over the proton polarizations in Eq.~\eqref{eq:StLR}.

Integrating Eq.~\eqref{eq:kineqLR} over $p_\mathrm{L}$ and $p_\mathrm{R}$, accounting for Eq.~\eqref{eq:StLR}, and multiplying the result by the factor $V/(2\pi)^{3}$, we obtain the kinetic equations,
\begin{align}\label{eq:kinNLR}
  \frac{\mathrm{d}N_\mathrm{L}}{\mathrm{d}t} = &
  W
  \left(
    R\to L
  \right) -
  W
  \left(
    L\to R
  \right),
  \nonumber
  \\
  \frac{\mathrm{d}N_\mathrm{R}}{\mathrm{d}t}= &
  W
  \left(
    L\to R
  \right) - W
  \left(
    R\to L
  \right),
\end{align}
for the total occupation numbers of left and right electrons, defined according to the expression,
\begin{equation}
  N_\mathrm{L,R}(t) =
  V\int
  \frac{\mathrm{d}^3p_\mathrm{L,R}}{(2\pi)^{3}}
  f_\mathrm{L,R}(\mathbf{p}_\mathrm{L,R},t).
\end{equation}
The total probabilities of transitions in Eq.~\eqref{eq:kinNLR} have the form,
\begin{align}\label{eq:WLR}
  W
  \left(
    R\to L
  \right)
  & =
  \frac{V}{2(2\pi)^8}
  \int
  \mathrm{d}^3p_\mathrm{L} \mathrm{d}^3p_\mathrm{R}
  \mathrm{d}^3 k_1 \mathrm{d}^3 k_2
  \frac{|\mathcal{M}_{R\to L}|^{2}}{\mathcal{E}_1\mathcal{E}_2}
  \delta^{4}
  \left(
    p_\mathrm{L}+k_2-p_\mathrm{R}-k_1
  \right)
  \notag
  \\
  & \times
  \left(
    1-f_\mathrm{L}
  \right)
  f_\mathrm{R}
  \left[
    1-f_{p}(\mathbf{k}_2)
  \right]
  f_{p}(\mathbf{k}_1),
  \nonumber
  \\
  W
  \left(
    L\to R
  \right)
  & =
  \frac{V}{2(2\pi)^8}
  \int
  \mathrm{d}^3p_\mathrm{L} \mathrm{d}^3p_\mathrm{R}
  \mathrm{d}^3 k_1 \mathrm{d}^3 k_2
  \frac{|\mathcal{M}_{L\to R}|^{2}}{\mathcal{E}_1\mathcal{E}_2}
  \delta^{4}
  \left(
    p_\mathrm{L}+k_1-p_\mathrm{R}-k_2
  \right)
  \notag
  \\
  & \times
  f_\mathrm{L}
  \left(
    1-f_\mathrm{R}
  \right)
  \left[
    1-f_{p}(\mathbf{k}_2)
  \right]
  f_{p}(\mathbf{k}_1).
\end{align}
In the first approximation one can replace $f_\mathrm{L,R}$ in Eq.~\eqref{eq:WLR} by the equilibrium distribution functions of electrons $f_e(E-\mu_\mathrm{L,R})$,
which are used in Sec.~\eqref{sub:MAT}. In this case, one can see that Eqs.~\eqref{eq:kinNLR} and~\eqref{eq:WLR} coincide with Eqs.~\eqref{eq:kineqN}
and~\eqref{eq:Wew} respectively.

\section{Energy source providing the magnetic field growth\label{sec:ENSOR}}

In this Appendix we consider the mechanism of the transformation of the thermal energy of background matter fermions to the energy of the growing magnetic field.

Despite gases of fermions in NS are highly degenerate, they have a nonzero temperature. For example, at $t \sim 10^2\thinspace\text{yr}$ after the SN explosion, the temperature can reach $T \gtrsim 10^8\thinspace\text{K}$. In Ref.~\cite{DvoSem15c} we suggested that the growth of the magnetic field, predicted in Refs.~\cite{DvoSem15a,DvoSem15b}, can be provided by the transmission of the thermal energy of background fermions. To justify the possibility of this process, one should consider the energy conservation equation in magnetohydrodynamics (MHD)~\cite[pp.~226--227]{LanLif84}:
\begin{equation}\label{eq:enbalMHD}
  \frac{\partial}{\partial t}
  \left(
    \frac{\rho \mathbf{v}^2}{2} + \varepsilon_\mathrm{T} + \frac{B^2}{2}
  \right) =
  - (\nabla \cdot \bm{q}),
\end{equation}
where $\rho$ is the mass density of the NS matter, $\varepsilon_\mathrm{T}$ is the internal energy per unit volume, $\mathbf{v}$ is the velocity, and $\bm{q}$ is the density of the energy flux.

One can represent $\varepsilon_\mathrm{T}$ in the form $\varepsilon_\mathrm{T} = \varepsilon_0 + \delta \varepsilon_\mathrm{T}$, where $\varepsilon_0$ is the internal energy of the degenerate gas, which does not depend on temperature, and $\delta \varepsilon_\mathrm{T}$ is the thermal correction. In Ref.~\cite{DvoSem15c} we showed that the magnetic field can take the energy from $\delta \varepsilon_\mathrm{T}$. Moreover, in Ref.~\cite{DvoSem15c}, we found that
\begin{equation}\label{Beq}
  \delta \varepsilon_\mathrm{T} =
  \frac{B_\mathrm{eq}^2}{2} = 
  \left[
    \frac{M_\mathrm{N} (p_{\mathrm{F}_n} + p_{\mathrm{F}_p})}{2} +\mu_e^2
  \right]
  \frac{T^2}{2},
\end{equation}
where $M_\mathrm{N}$ is the nucleon mass and $p_{\mathrm{F}_{p,n}}$ are the Fermi momenta of protons and neutrons.

Integrating Eq.~\eqref{eq:enbalMHD} over the NS volume $V$, assuming that $\bm{q} = 0$ at the NS surface, and accounting for Eq.~\eqref{Beq}, we get the conservation law,
\begin{equation}\label{eq:conslaw}
  \frac{\mathrm{d}}{\mathrm{d}t}
  \left(
    \delta \varepsilon_\mathrm{T} + \rho_\mathrm{B}
  \right) = 0,
  \quad
  \rho_\mathrm{B} = \frac{1}{2V} \int B^2 \mathrm{d}^3 x.
\end{equation}
Eq.~\eqref{eq:conslaw} shows that the growth of the magnetic energy happens owing to decreasing of the thermal correction to the internal energy. In Eq.~\eqref{eq:conslaw} we take into account that $\dot{\varepsilon}_0 = 0$.

Using Eq.~\eqref{eq:conslaw}, one can also substantiate the quenching of the parameter $\Pi$ in Eq.~\eqref{eq:Piq}. If we neglect the NS cooling due to the neutrino emission, then, integrating Eq.~\eqref{eq:conslaw} with the proper initial condition, one gets that $B^2 + B_\mathrm{eq}^2(T) = B_\mathrm{eq}^2(T_0)$, where we assume that $B_\mathrm{eq}(T_0) \gg B_0$ in a typical NS.
Thus one gets that the temperature of the NS matter will depend on the growing magnetic field as $T^2 = T_0^2[1 -B^2/{B}_\mathrm{eq}^2(T_0)]$, where we use Eq.~\eqref{Beq}. Then, accounting for the temperature dependence of the conductivity $\sigma_\mathrm{cond} \sim 1/T^2$ found in Ref.~\cite{Kel73}, one obtains that the conductivity becomes dependent on the growing magnetic field,
\begin{equation}\label{TB}
  \sigma_\mathrm{cond} \to
  \sigma_\mathrm{cond}
  \left[
    1 - \frac{B^2}{B_\mathrm{eq}^2(T_0)}
  \right]^{-1}.
\end{equation}
Note that, it is sufficient to take into account this dependence only in the terms in Eqs.~\eqref{eq:heq}-\eqref{eq:mu5eq} which contain $\mu_5+V_5$ since it is these terms which are responsible for the magnetic field instability. In Ref.~\cite{YakPet04} it was shown that, at $\sim 10^2\,\text{yr}$ after a SN explosion, mainly the neutrino emission in modified Urca-processes contribute to the NS cooling. If one accounts for this channel of the NS cooling as well, we should replace ${B}_\mathrm{eq}^2(T_0) \to {B}_\mathrm{eq}^2(T)$ in Eq.~\eqref{TB}. This modification of Eqs.~\eqref{eq:heq}-\eqref{eq:mu5eq} is equivalent to Eq.~\eqref{eq:Piq}. Therefore, the magnetic field amplification, predicted in our model, takes into account the NS cooling because of the neutrino emission.

It should be noted that we assume that $\bm{q}=0$ while deriving the conservation law in Eq.~\eqref{eq:conslaw}. This assumption is valid if we neglect the photon emission from the NS surface. Thus, if we take that $\bm{q}\neq 0$, i.e. the photon emission from the NS surface is present, then a part of the initial thermal energy will be spent in this channel of the NS cooling. Hence, the magnetic field strength $B_\mathrm{max}$, obtained in Sec.~\ref{sec:EVOL}, will be slightly overestimated. However, as shown in Ref.~\cite{YakPet04}, the photon emission from the NS surface is a subsidiary (compared to the neutrino emission, which, as was mentioned above, is taken into account exactly in our model) channel of the NS cooling in the time interval $10^2\,\text{yr}\lesssim t \lesssim 10^6\,\text{yr}$ used in Sec.~\ref{sec:EVOL}. Hence, if one derives the analogue of the conservation law in Eq.~\eqref{eq:conslaw} supposing that $\bm{q}\neq 0$, it will not lead to a significant deviation of $B_\mathrm{max}$ compared to Sec.~\ref{sec:EVOL}.

It is interesting to mention that, despite NS cools down because of the magnetic field growth, the second law of thermodynamics is not violated. This fact can be verified using the heat transfer equation in MHD, which results from 
Eq.~\eqref{eq:enbalMHD}, in the form~\cite[p.~229]{LanLif84},
\begin{equation}\label{eq:entropeq}
  \rho T
  \left[
    \frac{\partial s}{\partial t} +
    \left(
      \mathbf{v} \nabla
    \right) s
  \right] =
  \kappa \nabla^2 T +
  \frac{1}{\sigma_\mathrm{cond}}
  \left(
    \nabla \times \mathbf{B}
  \right)^2,
\end{equation}
where $s$ is the the entropy per unit mass and $\kappa$ is the thermal conductivity coefficient. Note that in Eq.~\eqref{eq:entropeq} we neglect the contribution of the viscosity tensor. Integrating Eq.~\eqref{eq:entropeq} over the NS volume, one gets the time evolution of the total entropy $S$~\cite[p.~195]{LanLif87},
\begin{equation}\label{eq:entropevol}
  \frac{\mathrm{d} S}{\mathrm{d}t} =
  \int \kappa \frac{(\nabla T)^2}{T^2} \mathrm{d}^3 x +
  \int \frac{
  \left(
    \nabla \times \mathbf{B}
  \right)^2}{T\sigma_\mathrm{cond}}
  \mathrm{d}^3 x,
  \quad
  S = \int \rho s \mathrm{d}^3 x.
\end{equation}
One can see in Eq.~\eqref{eq:entropevol} that $\dot{S} > 0$, i.e. the second law of thermodynamics is not violated.

At the end of this Appendix we mention that the magnetic cooling is well known in science and technology. Firstly, we mention the radiative cooling of electrons in a strong magnetic field with the formation of the two levels system of particles with opposite spins, which was described in Ref.~\cite{TerBagDor68}. Secondly, we recall the magnetocaloric effect, which has numerous technological applications~\cite{Yu03}.

\end{document}